\documentstyle[preprint,aps,prb]{revtex}
\begin{document}

\draft

\title{
DMRG study of the spin gap in a one dimensional Hubbard model: 
effect of the distant transfer and exchange coupling
}

\author{Ryotaro Arita, Kazuhiko Kuroki, Hideo Aoki}
\address{Department of Physics, University of Tokyo, Hongo,
Tokyo 113, Japan}
\author{Michele Fabrizio}
\address{Istituto Nazionale di Fisica della Materia, INFM, 
and International School for Advanced Studies, 
Via Beirut 2-4, 34013 Trieste, Italy}
\date{\today}

\maketitle

\begin{abstract}
The spin gap of a one-dimensional repulsive Hubbard model 
is numerically calculated with the 
density matrix renormalization group, with a special emphasis on 
the effect of a next-nearest neighbor hopping $(t')$ and 
the nearest-neighbor ferromagnetic exchange $(J)$ interaction. 
At half-filling, a significant spin gap
opens if $|t'| \simeq |t|$ and $J=0$,
in agreement with the weak coupling theory, while the gap is strongly
suppressed by the introduction of $J$.
On the other hand, the quarter-filled system has very small 
spin gaps regardless of the values of $t'$ and $J$. 
Implications for the 
CuO$_2$ chain in Sr$_{14}$Cu$_{24}$O$_{41}$ and related materials 
are discussed.

\end{abstract}

\medskip


\narrowtext

\newpage
For the past several years, the spin gap in strongly correlated 
one dimensional systems has been a focus of attention.
Intensive studies have been done on ladders
\cite{DagRice,Fabrizio,Schulz,Balents,Noack,Kuroki,Kimura} 
and systems with dimerized\cite{Imada} or frustrated\cite{Ogata} spin-spin 
interactions. 
Interests arose, since some of these studies have revealed that
the existence of a spin gap can lead to a dominance of
pairing correlations even in purely repulsive models.  

More recently, one of the present 
authors studied a one dimensional (1D) Hubbard model 
($t$-$t'$-$U$ model) that contains
next-nearest neighbor (NNN) hoppings in addition to nearest-neighbor (NN)
ones. Making use of a weak-coupling renormalization, 
he has predicted the existence of a spin gap 
in a certain parameter regime.\cite{Fabrizio2}  
A notable observation there is a prediction of 
a {\it superconductor-insulator} 
transition occurring at some 
finite Hubbard interaction $U=U_C$ 
at half-filling, for which three of the present authors   
subsequently have identified such a transition at a finite $U$ 
with quantum Monte Carlo (QMC) 
and density matrix renormalization (DMRG) methods.\cite{KuroAri}

The 1D $t$-$t'$-$U$ model is thus 
theoretically interesting, where the system has both a 
frustration (arising from the interference between $t$ and $t'$) 
and a character of a ladder as mentioned below.  
The model with $|t|\sim |t'|$ (where interesting 
physics is expected) may also 
have some relevance to real materials such as the cuprates 
that contain 1D chains.  
Specifically, a recently discovered superconductor, 
Sr$_{14-x}$Ca$_{x}$Cu$_{24}$O$_{41+\delta}$\cite{Uehara}, and related 
materials possess layers consisting of 1D chains alternating with 
layers of two-leg ladders. 
In the chain, the NN (Cu-Cu) transfer $(t)$ should be reduced 
due to the 90$^{\circ}$ Cu-O-Cu bond angle, just as in 
the inter-ladder hopping (estimated to be 
$\sim 0.1$ eV\cite{Rice}) in the ladder layers.
The NNN transfer $(t')$ in the chain is in turn equivalent to
the NNN transfer (estimated to be 
$\sim 0.1$ eV\cite{Hybertsen}) in the CuO$_2$ plane in the 
2D cuprates. We then end up with $|t|\sim |t'|$ for the chain.  
In fact, recent band calculation for Sr$_{14}$Cu$_{24}$O$_{41}$
shows that $t=-0.09$ eV and $t'=-0.17$ eV.\cite{Arai}

To be more precise, the effect of small $t$ not only gives
rise to a large ratio of $|t'/t|$, but also results in a net ferromagnetic 
NN exchange in real materials. Namely, the real materials has 
a ferromagnetic exchange interaction between NN Cu sites due to 
the exchange between Cu3d and O2p orbitals\cite{Braden,Mizuno} 
and the Hund's coupling between the O$p_x$ and O$p_y$ orbitals.\cite{Rice}
Since the antiferromagnetic NN interaction $(\sim 4t^2/U)$ is small,
the net NN interaction turns out to be ferromagnetic with an order of 
few hundred K in La$_6$Ca$_8$Cu$_{24}$O$_{41}$.\cite{Mizuno}
Such an ferromagnetic interaction is not included in the $t$-$t'$-$U$ model 
since the oxygen orbitals are not explicitly considered. 
In order to take this effect into account, we should add a 
NN ferromagnetic exchange $J$ to the $t$-$t'$-$U$ model to have 
the `$t$-$t'$-$U$-$J$' model.

In Sr$_{14}$Cu$_{24}$O$_{41}$, for which the experiments suggest that 
the average Cu valence in 1D chains is around +2.5,
a spin gap of about 140K is observed.\cite{Carter,Matsuda,Motoyama,Tsuji} 
The origin of the spin gap is yet to be established theoretically.
On the other hand, in La$_6$Ca$_8$Cu$_{24}$O$_{41}$, where 
the average Cu valence is +2, such a spin gap is not observed
in the 1D chain, and the system behaves as nearly free spins.
Since the weak-coupling theory predicts the existence of the spin gap
in the $t$-$t'$-$U$ model in a certain parameter regime, 
it would be intriguing to see if the experimental observations can be 
explained by the effect of the sizeable $|t'/t|$.
In La$_6$Ca$_8$Cu$_{24}$O$_{41}$, to be more precise, 
the magnetic susceptibility can be fitted with the Curie-Weiss law with
a Curie temperature of $\sim 10K$\cite{Carter,Motoyama}, which
is much smaller than the NN ferromagnetic exchange mentioned above.
It would also be interesting to investigate whether
such an observation can be understood within the interplay
of $t'$ and $J$.

The purpose of the present paper is to evaluate the spin gap
in the $t$-$t'$-$U$ and $t$-$t'$-$U$-$J$ models using DMRG, 
and to compare the results with the 
experimental observations for the cuprates.  
Since the Cu valence +2 and +2.5 corresponds to the band 
filling $n=1$ (half-filling) and $n=0.5$ (quarter-filling), respectively,
we focus on these fillings.

Let us start with the $t$-$t'$-$U$ model.
The Hamiltonian is given,
in standard notations, as
\begin{eqnarray*}
{\cal H}&=&-t\sum_{i \sigma}
(c_{i\sigma}^\dagger c_{i+1\sigma}+{\rm h.c.})
-t'\sum_{i \sigma}
(c_{i\sigma}^\dagger c_{i+2\sigma}+{\rm h.c.})\\
&&+U\sum_i n_{i\uparrow}n_{i\downarrow} ,
\end{eqnarray*}
where $t$, ($t'$) are NN (NNN) hoppings and $U$ is the 
Hubbard repulsion. The sign of $t$ does not change the physics 
because it can be changed by a gauge transformation.
Thus, we take $t>0$ without a loss of generality.
On the other hand, the sign of $t'$ is crucial when the 
system deviates from half-filling, at least in the weak-coupling theory.
Here we take $t'<0$ in accord with the cuprates. 

Let us first recapitulate the weak-coupling result.\cite{Fabrizio2}
When the system is mapped to 
the Tomonaga-Luttinger (TL) model, 
the system with a small enough  $|t'/t|$ reduces to 
the $t'=0$ case, 
since they are essentially the same in the vicinity of the 
Fermi level. 
For a substantial $|t'|$, on the other hand, 
the Fermi level intersects the one-electron 
band at four $k$-points 
($\pm k_F^1$ and $\pm k_F^2$). 
Namely, there are two right-moving and 
two left-moving branches in the terminology of the 
TL model.  
When the Umklapp processes are absent, 
the situation is, as far as the weak-coupling picture is concerned, 
essentially identical to the two-leg 
Hubbard ladder, where a spin gap opens when the two 
Fermi velocities are not too different.\cite{Balents} 
Thus, a spin gap
opens in the present model as well for a certain range of the ratio of the 
Fermi velocities, i.e., for $|t'|>t_c'(n)$. 

Bearing this in mind, 
let us now turn to the numerical calculation of the spin gap
performed by DMRG.
We take several typical values, namely, $t'=-0.55, -0.8,$ and $-2.0$ 
(hereafter $t=1$ is taken as a unit of the energy).
As for the value of $U$, we take $U=8$ for $t'=-0.55$ and $t'=-0.8$, 
and $U=16$ for $t'=-2$.  
We have taken such large 
values of $U$ because otherwise 
the spin gap strongly fluctuates with the number of sites. 
The fluctuation becomes worse for large $|t'|$, 
so that we take $U=16$ for $t'=-2$. 
Fortunately, large $U$ is of physical interest, 
since we can probe the strongly correlated systems 
such as the cuprates. 

Generally, the spin gap is much smaller than
the charge gap, so an accurate calculation is required.
The calculation has been done for system sizes up to
$L=56$ sites for $t'=-2$, and $L=48$ sites for $t'=-0.55$ or $-0.8$.
We kept up to 300 states per block, and 
stored the transformation matrix to construct the 
initial state for each superblock diagonalization.
\cite{accDMRG}
After each diagonalization, 
the accuracy of wave function is improved 
by the inverse iteration and
conjugate-gradient optimization, where 
the truncation error is $\sim O(10^{-5})$.

The spin gap of a system with $N$ up-spin and $N$ down-spin 
electrons is calculated by the equation,
\[
\Delta_S=E(N+1, N-1)-E(N, N),
\]
where $E(N_\uparrow, N_\downarrow)$ is the ground state energy of
a system with $N_\uparrow$ up spin and $N_\downarrow$ down spin electrons.

We first look at the spin gap for $t'=-0.8$ with $U=8$ 
in Fig. \ref{sgap08}, which plots $\Delta_S$ 
against the inverse of the system size, $1/L$. 
By least-square fitting the 
result with second order polynomials in $(1/L)$, 
we evaluate the spin gap in the 
thermodynamic limit.  
For half filling it is estimated to be $\Delta_S=0.16$.  
At quarter filling, the spin gap becomes very small, 
but is still finite as far as the present fitting is concerned.
For $t'=-0.8$, the Fermi level 
intersects the one-electron band (which has double minima) 
at four $k$-points both for $n=1$ and 
$n=0.5$, where the two Fermi velocities have comparable magnitudes.  
Hence our result is consistent with the weak-coupling theory, 
which is nontrivial since $U$ here is quite large.

Next we turn to the spin gap for a smaller $t'=-0.55$.  
For this value of $t'$, 
the Fermi level at half-filling 
still intersects the one-electron 
band at four points, but the inner pair 
is close to the local maximum ($k=0$), where the Fermi velocity is 
very small. According to the weak-coupling theory, the spin gap does not
open if the ratio of the two Fermi velocities is too small (or large).
Our DMRG result for $U=8$ 
in Fig.\ref{sgap55} 
indeed shows that the spin gap closes or becomes
negligibly small.  

At quarter-filling 
the system with $t'=-0.55$ and $U=8$ turns out to 
be {\it ferromagnetic}, 
as confirmed by DMRG calculation of the total energy and 
the total spin.
This is in accord with previous studies\cite{Pieri,Daul,comment}
showing that the ground state of the $t$-$t'$-$U$ model is
fully spin-polarized for sufficiently large $U$, small (but finite) $t'$,
and fillings away from half-filling.

Finally, we show the spin gap for a larger $t'=-2$ with 
$U=16$ in Fig. \ref{sgap2}. 
To further reduce the finite size effect,
we choose the system size of 
$4\times$ integer electrons.
At half-filling, the spin gap in the present fitting 
opens in agreement with the 
weak-coupling theory, but it is much 
smaller than that for $t'=-0.8$.
For the half-filling, 
a comparison with a DMRG result for the $J$-$J'$ Heisenberg model
(Heisenberg model with NNN antiferromagnetic exchange)
by White and Affleck\cite{White3} may be interesting.
According to their result, the
magnitude of the spin gap 
as a function of $J'/J$ 
takes its maximum at $J'/J \simeq 0.6$,
and becomes much smaller for $J'/J > 2$.
If we assume $J=4t^2/U$ and $J'=4t'^2/U$, 
we can then draw a consistent picture 
that the magnitude of the spin gap of the $t$-$t'$-$U$ model
takes its maximum at $t' \simeq -0.8$, and is very small for 
$|t'|> \sqrt 2 t$. 

At quarter-filling for $t'=-2$, the magnitude of the spin gap 
further decreases to become negligibly small 
in the thermodynamic limit (unless the curve upturns around $1/L \sim 0$). 
If the spin gap does exactly vanish, 
this cannot be understood within the weak-coupling theory
because the Fermi level intersects the band at four points, and
the ratio of the Fermi velocities is not large. The intraband umklapp 
processes are known to suppress the spin gap in the Hubbard ladders,
\cite{Balents,Noack,Kuroki} but for $t'=-2$ and $n=0.5$ in the present model,
processes corresponding to the intraband umklapp do not exist.
Then, the disappearance of the spin gap in the region of large $U$, if any, 
would be beyond the scope of the weak-coupling theory.

To summarize the above results,
a relatively large spin gap opens at half-filling for $|t'|\simeq t$,
while at quarter filling, either the magnitude of the spin gap becomes
smaller, or the ground state becomes ferromagnetic. 
The suppression of the 
spin gap for $n=0.5$ may be reminiscent of the 
results of the $t$-$J$-$J'$ model obtained
by Ogata {\it et al}, where the spin gap at half-filling 
is destroyed by hole doping.\cite{Ogata} We stress that this is not 
trivial since the effective Hamiltonian of the doped $t$-$t'$-$U$ model
in the large $U$ limit is not the $t$-$J$-$J'$, but the $t$-$t'$-$J$-$J'$
model, where the NNN hopping may play an crucial role.

We move on to the result for the $t$-$t'$-$U$-$J$ model 
with the Hamiltonian
\begin{eqnarray*}
{\cal H}&=&-t\sum_{i \sigma}
(c_{i\sigma}^\dagger c_{i+1\sigma}+{\rm h.c.})
-t'\sum_{i \sigma}
(c_{i\sigma}^\dagger c_{i+2\sigma}+{\rm h.c.})\\
&&+U\sum_i n_{i\uparrow}n_{i\downarrow}
-J\sum_{i}\left( {\bf S}_i \cdot {\bf S}_{i+1} +
\frac{1}{4}n_i n_j \right),
\end{eqnarray*}
where $J>0$ is the ferromagnetic exchange. 
In Fig. \ref{J}(a), we show the result for the spin gap 
when $J=0.5$ or $1.0$ is present for
$t'=-0.8$, $U=8$ at half-filling 
to compare with the result for $J=0$.  
We can see that the 
even the spin gap for $t'=-0.8$, which has a relatively large value in the 
absence of $J$, is almost completely suppressed by $J=0.5$.  
Intuitively this is because the spin-gap behavior arising 
from the effect of $t'$ is `canceled out' 
with the effect of $J$, which suppresses the spin fluctuations.
Since the spin gap takes its maximum around $t'=-0.8$ for $J=0$, 
the present result suggests that $J=0.5$ is enough to suppress 
the spin gap at other values of $t'$.
At quarter filling, where the spin gap is extremely small
already at $J=0$, the introduction of $J$ further destroys the gap
as shown in Fig.\ref{J}(b) for the case of 
$J=0.5$ or $1.0$ with $t'=-2$, $U=16$.  

In both the half- and quarter-filled cases, large enough $J$ should
produce a ferromagnetic ground state. For the present values of $J$
and the system sizes, we do not have a ferromagnetic ground state,
as seen from the fact that the spin gap (at each system size) is finite.
However, we do have a possibility that a ferromagnetic ground state
is realized in the thermodynamic limit, especially in the cases where
the extrapolated spin gap is negative, as in $n=1$ and $J=1$, or
in all quarter filled cases.

Let us finally comment on the implications of the present 
results on the experimental observations.  
In La$_6$Ca$_8$Cu$_{24}$O$_{41}$, the magnetic susceptibility in the 
1D chains does not indicate a spin gap. 
Our numerical results suggest that this can be explained within the 
half-filled $t$-$t'$-$U$-$J$ model. 
Namely, the net NN ferromagnetic interaction $J_{\rm net}$
evaluated for the 1D chain in the cuprates is about few Hundred K,
\cite{Mizuno} which corresponds to $J(\sim J_{\rm net}+4t^2/U)\sim 0.5t$,
if we adopt the band calculation result  
of $t\sim 0.1$ eV for Sr$_{14}$Cu$_{14}$O$_{41}$,\cite{Arai}
and assume $U\sim O(10t)$.
As mentioned above, $J\sim 0.5t$ is enough to suppress the spin gap
for the cases investigated at half-filling, and might even produce 
ferromagnetism. 
Thus, the situation corresponding to the real materials is subtle, 
and the ground state may become weakly ferromagnetic depending on the 
details of the parameter values, which may also be consistent
with the small Curie temperature of $\sim 10$ K observed experimentally.
\cite{Carter,Motoyama}
The present results suggest that
the weakly ferromagnetic behavior of the susceptiblity is not due to 
the interactions being extremely weak, but due to the `cancellation effect'
between $t'$ and $J$.

On the other hand, the spin gap is observed for the 1D chain in 
Sr$_{14}$Cu$_{24}$O$_{41}$ with $\Delta_S \sim 140$K, 
which corresponds to $\sim 0.1t$ (assuming $t\sim 0.1$ eV) in
the $t$-$t'$-$U$ model. 
All our numerical results at quarter filling 
show that the spin gap is much smaller than $0.1t$, if any. 
Thus, we may conclude that the 
experimentally observed spin gap cannot be explained within 
$t$-$t'$-$U$ or $t$-$t'$-$U$-$J$ models.
Then, some other ingredients, e.g., 
a coupling with the lattice (which may be dimerized), should be 
necessary. 
Even in that case, there still remains a possibility that
$t'$ is playing an important role in opening the spin gap. 
Further investigation will be required to clarify this point.

R.A. wishes to thank T. Nishino, N. Shibata,
Y. Nishiyama for helpful discussions on 
the DMRG method. We also thank K. Kusakabe and T. Kimura
for valuable discussions. 
Numerical calculations were done on FACOM VPP 500/40 at the Supercomputer 
Center, Institute for Solid State Physics, University of Tokyo, and 
HITAC S3800/280 at the Computer Center of the University of Tokyo.
This work was also supported in part by Grant-in-Aid for Scientific
Research from the Ministry of Education of Japan. 


\begin{figure}
\caption{DMRG result for the 
spin gap, $\Delta_S$, 
in the $t$-$t'$-$U$ model 
plotted against the inverse system size $1/L$ 
for $t'=-0.8$, $U=8$ with the band filling 
$n=1$ or $n=0.5$. 
Solid curves are least-square fits to second
order polynomials in $1/L$. 
}
\label{sgap08}
\end{figure}

\begin{figure}
\caption{A plot similar to Fig. 1 for $t'=-0.55$, $U=8$.}
\label{sgap55}
\end{figure}

\begin{figure}
\caption{A plot similar to Fig. 1 for
for $t'=-2$, $U=16$.}
\label{sgap2}
\end{figure}

\begin{figure}
\caption{A plot similar to Fig. 1 
for the $t$-$t'$-$U$-$J$ model 
for (a) $n=1$, $t'=-0.8$, $U=8$ and 
(b)$n=0.5$, $t'=-2$, $U=16$ with $J=0.0$, $J=0.5$ and $1.0$}
\label{J}
\end{figure}


\begin{references}
\bibitem{comment}In ref.\onlinecite{Daul}, 
the ferromagnetic phase boundary $U_c(t')$ 
was determined for $n\leq 1$ at three points 
$t'=-0.2$, $-0.8$, and $-2.0$.
We have more closely scanned the region $-0.8\leq t' \leq -0.2$ 
at quarter-filling with DMRG, to find that, 
although $U_c(t'=-0.2)=8$ and $U_c(t'=-0.8)=20$ 
in agreement with ref.\onlinecite{Daul}, 
$U_c(t')$ does not monotonically increases with $|t'|$ 
but $U_c$ has in fact its minimum 
($U_c=4$) around $t'=-0.4$.  
Hence $U_c(t'=-0.55) \simeq 5$, so that the ground state at $t'=-0.55$
and $U=8$ is fully polarized.
\end{references}
\end{document}